\documentclass[twocolumn]{aastex63}
\usepackage{graphics,epsf}
\usepackage{amsmath}                
\usepackage{amsfonts}               
\usepackage{amssymb}                
\usepackage{epsfig}                 
\usepackage{appendix}
\usepackage{graphicx}
\usepackage{float}
\usepackage{color}
\usepackage{multirow}
\usepackage{colortbl}
\usepackage[para,online,flushleft]{threeparttable}

\newcommand{\erg}{{~\rm erg}}
\newcommand{\yr}{{~\rm yr}}

\newcommand{\Gyr}{{~\rm Gyr}}

\newcommand{\AU}{{~\rm AU}}

\newcommand{\days}{{~\rm days}}

\begin{document}

\title{A red giant branch common envelope evolution scenario for the exoplanet WD~1856~b}


\author{Ariel Merlov}
\affiliation{Department of Physics, Technion – Israel Institute of Technology, Haifa 3200003, Israel; ealealbh@gmail.com; soker@physics.technion.ac.il}
  
\author{Ealeal Bear}
\affiliation{Department of Physics, Technion – Israel Institute of Technology, Haifa 3200003, Israel; ealealbh@gmail.com; soker@physics.technion.ac.il}

\author[0000-0003-0375-8987]{Noam Soker}
\affiliation{Department of Physics, Technion – Israel Institute of Technology, Haifa 3200003, Israel; ealealbh@gmail.com; soker@physics.technion.ac.il}
\affiliation{Guangdong Technion Israel Institute of Technology, Guangdong Province, Shantou 515069, China}

\begin{abstract}
We propose a common envelope evolution (CEE) scenario where a red giant branch (RGB) star engulfs a planet during its core helium flash to explain the puzzling system WD~1856+534 where a planet orbits a white dwarf (WD) of mass $M_{\rm WD} \simeq 0.52 M_\odot$ with an orbital period of $P_{\rm orb}=1.4 \days$. At the heart of the scenario is the recently proposed assumption that the vigorous convection that core helium flash of RGB stars drive in the core excite waves that propagate and deposit their energy in the envelope. Using the \textsc{binary-mesa} stellar evolution code we show that this energy deposition substantially reduces the binding energy of the envelope and causes its expansion. We propose that in some cases RGB stars might engulf massive planets of $\ga 0.01 M_\odot$ during their core helium flash phase, and that the planet can unbind most of the mass of the bloated envelope. We show that there is a large range of initial orbital radii for which this scenario might take place under our assumptions. This scenario is relevant to other systems of close sub-stellar objects orbiting white dwarfs,  like the brown dwarf-WD system ZTFJ003855.0+203025.5.
\end{abstract}


\section{Introduction} 
\label{sec:intro}

\cite{Vanderburgetal2020}
report the detection of a planet orbiting a white dwarf (WD~1856+534; TIC~267574918) with a period of $P_{\rm orb}=1.4 \days$ and an orbital separation of $a \simeq 0.02 \AU$ (also \citealt{Alonsoetal2021}). They further argue that this relatively long orbital period of the planet candidate makes a common-envelope evolution (CEE) origin of the system less likely than a process where a third body scatter the planet to this orbit. They find the present mass of the WD as $M_{\rm WD} = 0.518 \pm 0.055 M_\odot$ and its cooling age as $5.85 \pm 0.5 \Gyr$, implying that the progenitor mass should have been $\ga 1.1 M_\odot$. We will use for our study a stellar model with a zero age main sequence (ZAMS) mass of $M_{\rm ZAMS} =1.6 M_\odot$, but note that our scenario might work better for lower masses.
   
There were earlier claims for exoplanet candidates orbiting WDs (e.g., \citealt{Gansickeetal2019, Manseretal2019}). One earlier claim for a planet candidate around an horizontal branch star by \cite{Setiawanetal2010} was refuted by \cite{JonesJenkins2014}. \cite{Setiawanetal2010} claim for a planet with an orbital period of $16.2 \days$ orbiting a metal-poor horizontal branch star (for other refuted claims for planets around horizontal branch stars see, e.g., \citealt{Krzesinskietal2020}). That refuted system
had two extreme properties for a post-CEE surviving planet, a large semi-major axis of $\simeq 25 R_\odot$, and a large envelope mass of $\simeq 0.3 M_\odot$. \cite{Bearetal2011} proposed a speculative scenario where a metal-poor red giant branch (RGB) star suffers a rapid expansion during its core helium flash and engulfs a planet (see criticism by \citealt{Passyetal2012}). The very extended RGB envelope has a low binding energy and the planet survives the CEE by ejecting the envelope (section \ref{sec:Scenario}). 
 
There are many studies of planets influencing RGB and AGB stars (e.g., \citealt{Soker1998, NelemansTauris1998, SiessLivio1999a, NordhausBlackman2006,  Carlbergetal2009, Kunitomoetal2011, MustillVillaver2012, NordhausSpiegel2013, Villaveretal2014, AguileraGomezetal2016, Geieretal2016, Guoetal2016, Priviteraetal2016, Raoetal2018, Schaffenrothetal2019, Jimenezetal2020, Krameretal2020}). It seems that when an RGB star engulfs a planet the planet has a very low probability to survive the CEE because it cannot release enough orbital energy to unbind the envelope before it suffers destruction near the RGB core. Extra energy deposition to the envelope just before the CEE lowers the envelope binding energy and might allow a massive planet of mass $M_{\rm p} \ga {\rm few} \times M_{\rm J}$ to survive the CEE, where $M_{\rm J}$ is Jupiter mass. 

In a new study \cite{Bearetal2021} propose that waves that the vigorous convection during the core helium flash excite, might cause the envelope of RGB stars to substantially expand within few years. 
We here use this expansion to propose (section \ref{sec:Scenario}) and examine (section \ref{sec:Engulfment}) a CEE scenario for the formation of the planet-WD system WD~1856+534.
There are other scenarios for the formation of the system WD~1856+534. One group of studies examine the formation of this system by the scattering-in of the planet to an orbit around the WD after the formation of the WD, either planet-planet scattering in a multiple-planets system \citep{Maldonadoetal2021}, or scattering-in by a secondary star (or a tertiary star) in the system, i.e., the Lidov-Kozai effect (e.g., \citealt{Vanderburgetal2020, MunozPetrovich2020, OConnoretal2021, Stephanetal2020}). 

The other group of studies attribute the system WD~1856+534 to a CEE. \cite{Lagosetal2021} present the motivation to consider a CEE, and propose that the CEE takes place on the AGB. For their scenario to work they need an extra energy source (in addition to the orbital energy of the planet) to remove the entire envelope. We, instead, consider the CEE to take place on the RGB. As well, they show that the planet survives the post-CEE against evaporation. We build on these parts of their study.
\cite{Chamandyetal2021} attribute the extra energy source to another planet in the system that entered the RGB or AGB envelope at an earlier phase, and deposited a large fraction of the envelope binding energy. Such a process influences the evolution of the planet that orbits further out and might help it to survive (e.g., \citealt{Bearetal2011, Lagosetal2021}).     
 
\section{The basic scenario and assumptions}
\label{sec:Scenario}

The unique ingredient of the scenario that we deal with here for an RGB star to engulf an exoplanet(s) during its core helium flash is that during the core helium flash on the termination of the RGB phase, the vigorous helium burning in the core leads to the deposition of energy in the envelope. This energy causes the envelope expansion. 

\cite{Bearetal2011} considered the energy source to be the ignition of hydrogen at the base of the hydrogen-rich envelope in metal poor stars. They based their speculative scenario on the results of \cite{Mocaketal2010} who calculated hydrogen ignition by the core helium flash, a process that releases $\approx 1 \times 10^{48} \erg$ of nuclear energy during the first year. 
\cite{Bearetal2011} manually added an energy of $E_{\rm in} = 8.5 \times 10^{46} \erg$ just above the hydrogen-burning shell in a time period of 7 years at an average power of $L_{\rm in} = 10^5 L_\odot$ and found the star to expand by a factor of about 4. 
We cannot apply this scenario to stars with solar metalicity or higher. 

We apply the scenario that \cite{Bearetal2021} propose where waves that the vigorous convection during the core helium flash excite propagate to the envelope and deposit their energy there. \cite{Bearetal2021} base their scenario on the results of \cite{QuataertShiode2012} and \cite{ShiodeQuataert2014} who study the propagation from the core to the envelope of waves that the vigorous core convection in pre-supernova massive stars excite. The waves deposit their energy in the envelope causing it to expand (e.g., \citealt{McleySoker2014, Fuller2017}). 

\begin{figure}
\includegraphics[trim=2.1cm 6.9cm 0.0cm 7.2cm ,clip, scale=0.52]{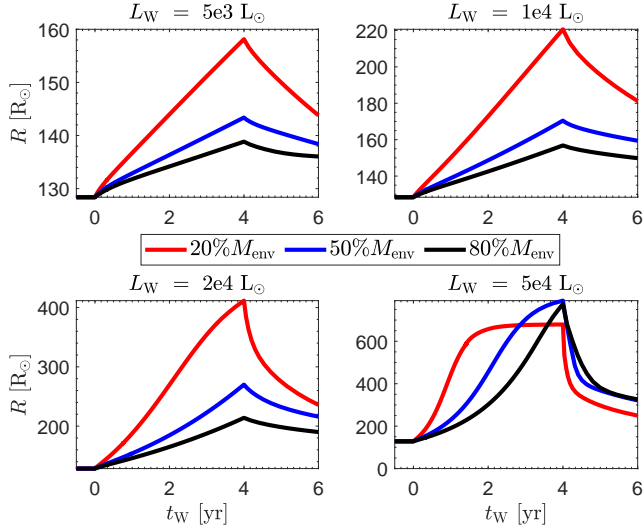}
\caption{The radius as function of time as a result of wave-energy deposition into the RGB envelope of a stellar model with initial mass of $M_{\rm ZAMS}= 1.6 M_\odot$. Each panel shows the results for one value of the waves power $L_{\rm W}$ as indicated, and for three cases according to the outer envelope mass into which the waves deposit their energy. In all cases the energy deposition time period lasts for four years. We set $t_{\rm W}=0$ at the beginning of energy deposition (from \citealt{Bearetal2021}). }
 \label{fig:R_vs_time}
 \end{figure}

For their model of $M_{\rm ZAMS}= 1.6 M_\odot$ that we use here, \cite{Bearetal2021} apply a formula from \cite{LecoanetQuataert2013} and find the total energy that the waves might deposit to the envelope to be $E_{\rm wave,0}  = 2.1 \times 10^{47} \erg =1.7 \times 10^6 L_\odot \yr$. They took a conservative approach and deposit less than this energy to the envelope at a constant luminosity $L_{\rm W}=\beta E_{\rm wave,0}/\Delta t_{\rm dep}=4.3 \times 10^5 \beta L_\odot $ during a time period of $\Delta t_{\rm dep}=4 \yr$ and with $\beta \ll 1$. Because of the uncertainty in the location in the envelope where the waves deposit their energy they examine three prescriptions. They deposit the wave energy to the envelope outer $\xi M_{\rm env}$ mass, with $\xi=80\%$, $\xi=50\%$, or $\xi=20\%$, and with a constant power per unit mass. The core and envelope mass when we deposit the wave energy are $M_{\rm core,b}=0.45 M_\odot$ and $M_{\rm env,b}=1.01 M_\odot$, respectively (the subscript `b' stands for just before energy deposition). 
In Fig. \ref{fig:R_vs_time}  we present the response of the envelope radius to wave-energy deposition for four values of wave power as \cite{Bearetal2021} present it.  

Based on this rapid expansion of the RGB star we propose the following scenario for the formation of the planet-WD system WD~1856+534. 
The rapid expansion during the core helium flash brought the RGB to engulf one or more of its exoplanets. The planet spirals-in in a time period of several years alongside the contraction of the RGB star. According to \cite{Vanderburgetal2020} the WD mass is 
$M_{\rm WD} = 0.518 \pm 0.055 M_\odot$ and the orbital separation is about $a=4 R_\odot$. For a stellar remnant mass of $0.52 M_\odot$ and an orbital separation of $a=4 R_\odot$ the planet of mass $M_{\rm p}$ releases an orbital energy of 
\begin{equation}
E_{\rm orb} = 2.5 \times 10^{45} 
\left( \frac{M_{\rm p}}{0.01 M_\odot} \right) \erg. 
\label{eq:Eorb}
\end{equation}
In our simulations (section \ref{subsec:Orbital}) we use a planet of mass $M_{\rm p} = 0.01M_\odot=10.5M_{\rm J}$.

The binding energy of the RGB envelope residing above mass coordinate $m=0.52M_\odot$ without wave energy deposition is $E_{\rm env,bind,b} (0.52) =1.2\times 10^{46} \erg$. 
We simulate the evolution of planets with two cases of wave energy deposition $(L_{\rm W}, \xi)=(2 \times 10^4 L_\odot, 20\%)$ and $(L_{\rm W}, \xi)=(5 \times 10^4 L_\odot, 80\%)$. The binding energy of the envelope that resides above mass coordinate $m=0.52 M_\odot$ at the end of wave energy deposition in the first case is $E_{\rm env,bind,20}(0.52) =6.4\times 10^{45} \erg$.  
The ratio $E_{\rm orb}/E_{\rm env,bind,20}(0.52) \simeq 0.4$ implies that the spiralling-in planet can unbind a large fraction of the envelope.

In the second case the energy of that envelope mass becomes positive, i.e. a negative binding energy of $E_{\rm env,bind,80} (0.52) =-6.2 \times 10^{45} \erg$. The envelope does not unbind itself despite its positive energy because the envelope ejection time, which is about the dynamical time of the extended envelope $\simeq 2 \yr$, is longer than the time that the envelope radiates this extra energy out, $\approx \vert E_{\rm env,bind,80} (0.52) \vert / L_{\xi=80} \simeq 1 \yr $, where $L_{\xi=80} \simeq 5 \times 10^4 L_\odot$ is the maximum luminosity that the star reaches at $t_{\rm W} = 4 \yr$ \citep{Bearetal2021}.
Nonetheless, we expect a highly enhanced mass loss during this phase, something that \textsc{mesa} does not include. The highly enhanced mass loss rate takes place after the planet already approaches the envelope because of tidal forces and spins the envelope up (before it even enters the envelope and after it enters the envelope). Because the planet is already falling towards the envelope and tidal forces are already large, the extra mass loss is not sufficient to prevent engulfment. 

There are other planet-induced effects that can enhance the mass loss rate. Excitation of p-waves by the planet (e.g., \citealt{Soker1993}) and the spinning-up of the envelope (e.g., \citealt{Soker1998AGB, NordhausBlackman2006})  can facilitate formation of dust that more efficiently couples the stellar radiation to wind and by that enhances the mass loss rate (e.g., \citealt{Soker1998AGB, GlanzPerets2018, Iaconietal2019}). We suggest that due to the rapid expansion during the core helium flash the planet manages to eject the envelope and survive. 

We attribute the same scenario for the formation of the system ZTFJ003855.0+203025.5 of a brown dwarf of mass $\simeq 0.059M_\odot$ orbiting a WD of mass $\simeq 0.5 M_\odot$ with a semi-major axis of $2.0 R_\odot$ as \cite{vanRoesteletal2021} reported recently. 

\section{Planet engulfment during the core helium flash}
\label{sec:Engulfment}
 
\subsection{Numerical setting}
\label{subsec:MESA}
We use \textsc{mesa-binary} version 10398 \citep{Paxtonetal2011,Paxtonetal2013,Paxtonetal2015,Paxtonetal2018,Paxtonetal2019}.
We divide our numerical simulations to two numerical phases: In all numerical phases in our binary inlist we follow the example of \textsc{mesa-binary} $star~plus~point~mass$. We set tidal for the binary system ($do~tidal~sync=.true.$).

In numerical phase A we follow the evolution of a $M_{\rm ZAMS} = 1.6 M_\odot$ star using the example of $1M~pre~ms~to~wd$, orbited by a planet of mass $M_{\rm p} = 0.01 M_\odot$. We treat the planet as a point mass. For each case of an initial orbital radius $a_0$ we find the time when the radius is maximal (this is consistent with the He flash) and we stop this numerical phase at 4 years before the maximal radius is achieved. In numerical phase B that lasts from $t_{\rm W} =0$ to $t_{\rm W}=4 \yr$, we manually insert energy (the wave energy) in the src folder in the run-star-extra.f in the subroutine: subroutine energy-routine file, when we set the pointer of $other~energy$ to true. As in \cite{Bearetal2021} we insert energy at a constant power $L_{\rm W}$ into the outer $\xi M_{\rm env}$ zone of the envelope. 
We analyse the influence of energy deposition on the radius of the star and on the orbital separation (radius).

\subsection{Orbital evolution}
\label{subsec:Orbital}

\begin{figure}
\includegraphics[trim=3.3cm 8.5cm 0.0cm 8.5cm ,clip, scale=0.62]{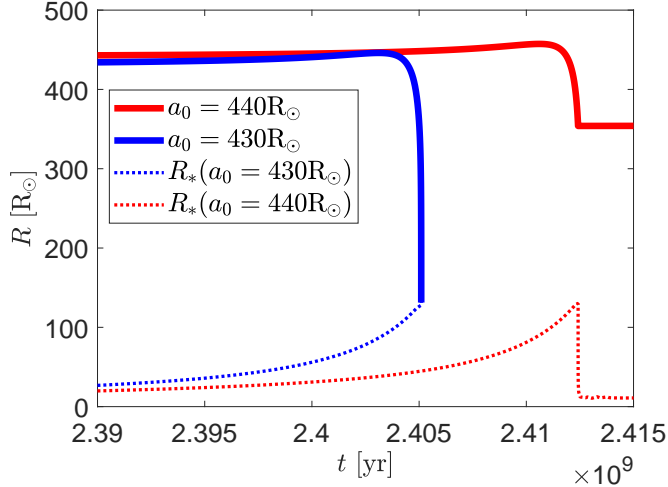}
\caption{The RGB radius (dotted line) and the orbital radius of the $M_{\rm p} = 0.01 M_\odot$ planet (thick solid line) as function of time at the end of the RGB evolution without wave energy deposition. The blue lines represent the case of an initial orbital radius (at ZAMS of the star) of $a_0=430 R_\odot$ for which the RGB star engulfs the planet, and the red lines represent the case of an initial orbital radius of $a_0=440 R_\odot$ for which the RGB star does not engulf the planet. }
\label{fig:NoWaveEnergy}
\end{figure}

In all simulations we take a planet of mass $M_{\rm p} = 0.01M_\odot=10.5M_{\rm J}$ and circular orbits. For each of the two wave energy deposition cases we search for the range of initial orbital radii (semi-major axes) $a_{\rm 0,min} \la a_{\rm 0,in} \la a_{\rm 0,max}$ for which the RGB star engulfs the planet during its rapid expansion following the core helium flash (but not before that). We first determine that for the planet to survive to the core helium flash its initial orbital radius should be $a_0 > a_{\rm 0,min}\simeq 440 R_\odot$.
In Fig. \ref{fig:NoWaveEnergy} we present the evolution on this boundary of engulfment without wave energy deposition, i.e., we present two cases with close initial orbital radii to each other where in one the RGB star engulfs the planet and in the other case that has a few percent larger initial radius the planet avoids engulfment. 

\begin{figure}
\includegraphics[trim=3.3cm 8.5cm 0.0cm 8.5cm ,clip, scale=0.62]{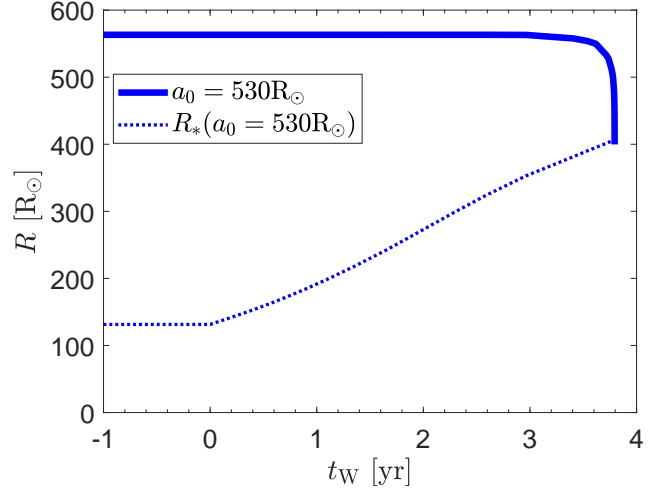} \\
\includegraphics[trim=3.3cm 8.5cm 0.0cm 8.5cm ,clip, scale=0.62]{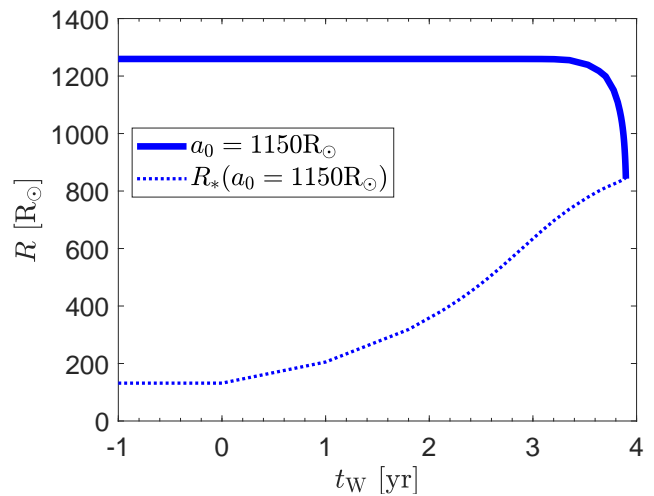}
\caption{The RGB radius (dotted line) and the orbital radius of the $M_{\rm p}=0.01 M_\odot$ planet (thick solid line) as function of time around the time of wave energy deposition to the envelope, $2.41 \times 10^9 \yr$. Both cases are for initial planetary orbits $a_0$ close to the upper boundary of planet engulfment. Because of mass loss the orbital radius increases by the time we deposit the wave energy. We start energy deposition at $t_{\rm W}=0$.
Upper panel: The case of $(L_{\rm W}, \xi)=(2 \times 10^4 L_\odot, 20\%)$, i.e., wave power of $L_{\rm W}=2 \times 10^4 L_\odot$ and energy deposition into the outer $\xi =20 \%$ mass of the envelope. 
Lower panel:  The case of $(L_{\rm W}, \xi)=(5 \times 10^4 L_\odot, 80\%)$.   }
\label{fig:WithWaveEnergy}
\end{figure}

For the cases of $(L_{\rm W}, \xi)=(2 \times 10^4 L_\odot, 20\%)$ and $(L_{\rm W}, \xi)=(5 \times 10^4 L_\odot, 80\%)$ we find the initial orbital radii for which the RGB star engulfs our planet during its core helium flash (the four years during which we deposit the wave energy) to be 
\begin{eqnarray}
\begin{aligned}
& 435 R_\odot \la a_{\rm 0,in,20}  \la 540 R_\odot \qquad {\rm and}
\\ 
& 435 R_\odot \la a_{\rm 0,in,80}  \la 1160 R_\odot, 
\label{eq:Separations}
\end{aligned}
\end{eqnarray}
respectively. The uncertainties in the values of the above boundaries that we find with \textsc{mesa} are $\simeq \pm 2 \%$ (not including uncertainties in some chosen parameters that we use in \textsc{mesa}). In Fig. \ref {fig:WithWaveEnergy} we present the evolution of the RGB radii and orbital separations during the period of the wave energy deposition for an initial orbital separation very close to the upper limit for planet engulfment.   

The main conclusions from our simulations that aim at explaining the planet-WD system WD~1856+534 are that  under our assumptions ($i$) the orbital energy that the planet releases is a significant fraction of the envelope binding energy after wave energy deposition, and ($ii$) there is a large range of initial planetary orbits for which the RGB engulfs the planet during the core helium flash.

\section{Summary}
\label{sec:summary}

We propose a scenario to explain the puzzling system WD~1856+534 where a planet orbits a WD of mass $M_{\rm WD} \simeq 0.52 M_\odot$ with an orbital period of $P_{\rm orb}=1.4 \days$ \citep{Vanderburgetal2020}. We chose the parameters of our numerical simulations, $M_{\rm ZAMS}=1.6 M_\odot$ and $M_{\rm p} = 0.01 M_\odot$, to comply with this system. We note though that the planet might be somewhat more massive (but still be a planet) and that the initial stellar mass can be as low as $M_{\rm ZAMS,min}=1.1 M_\odot$ \citep{Vanderburgetal2020}, both of which make our scenario more likely even. 

We base our study on the, yet to be tested, assumption of \cite{Bearetal2021} that the vigorous core convection during the core helium flash of RGB stars excite waves that propagate to the envelope and deposit their energy in the envelope, causing its expansion (Fig. \ref{fig:R_vs_time}) and substantially reducing its binding energy. It is sufficient that the energy that the waves carry during their few years activity is only $\simeq 5-10 \%$ of the possible wave energy that \cite{Bearetal2021} estimate from studies of massive stars \citep{LecoanetQuataert2013}.  

Our calculations under the above assumption that convection-induced waves cause RGB envelope expansion show that ($i$) an $M_{\rm p} \ga 0.01M_\odot$ planet that spirals-in inside the bloated RGB envelope can release  sufficient orbital energy to unbind a significant fraction of the loosely bound envelope, and ($ii$) there is a large range of initial planetary orbits for which the RGB engulfs the planet during the core helium flash (equation \ref{eq:Separations}).

As we mentioned in section \ref{sec:intro} some earlier studies noticed that an inner planet that enters the RGB (or AGB) envelope before the planet that eventually survives does, might remove envelope mass and allow the surviving planet to eject most of the envelope and survive. The presence of an inner planet or more can also increase the allowed parameter space for our proposed scenario.

We consider our proposed core helium flash wave energy scenario to be a promising explanation to the planet-WD system WD~1856+534 and similar systems of sub-stellar objects closely orbiting WDs, e.g., the brown dwarf-WD system ZTFJ003855.0+203025.5 that \cite{vanRoesteletal2021} recently analysed. 

\acknowledgments
We thank an anonymous referee for helpful suggestions. This research was supported by a grant from the Israel Science Foundation (769/20).  

\textbf{Data availability}

The data underlying this article will be shared on reasonable request to the corresponding author. 


\pagebreak

\end{document}